\documentclass[reqno]{amsart} \usepackage{amscd}
\usepackage{epsf}
\newtheorem{theorem}{Theorem}[section]
\newtheorem{proposition}[theorem]{Proposition}
\newtheorem{lemma}[theorem]{Lemma}

\theoremstyle{remark} 
\numberwithin{equation}{section}
\newcommand{\field}[1]{\ensuremath{\mathbb{#1}}}
\newcommand{\CC}{\field{C}}

\newcommand{\RR}{\field{R}}

\begin{document}
\title[Geometric prequantization of vortices]{Geometric Prequantization of the Moduli Space of the 
Vortex equations on a Riemann surface}
\author{Rukmini Dey}

\maketitle

\begin{abstract}
The moduli space of solutions to the vortex equations on a Riemann surface 
 are well known to have a symplectic (in fact K\"{a}hler) structure. We show 
this symplectic 
structure explictly and proceed to show a family of symplectic 
(in fact, K\"{a}hler) structures $\Omega_{\Psi_0}$ on the moduli space, 
parametrised by $\Psi_0$, a section of a line bundle
on the Riemann surface.  Next we show that corresponding to these there is a 
family of  prequantum line bundles ${\mathcal P}_{\Psi_0} $on the moduli space 
whose curvature is proportional to the symplectic forms $\Omega_{\Psi_0}$.  
\end{abstract}         

\section{Introduction}

Geometric 
prequantization  is a construction, if possible,  of a prequantum line bundle 
${\mathcal L}$ 
on a symplectic manifold, (${\mathcal M}, \Omega)$ whose curvature is proportional to the symplectic form.
The Hilbert space of the quantization is the space of  the square integrable sections of 
${\mathcal L}$.  
To every $f \in C^{\infty}({\mathcal M})$ we associate an operator acting on the Hilbert space, namely,  $\hat{f} = -i \hbar [X_f - \frac{i}{\hbar}\theta(X_f)] + f$ 
where $X_f$  is the vector field 
defined by $\Omega(X_f, \cdot) = - df$ and  $\theta$ is a symplectic 
potential correponding to $\Omega$. 
Then if $f_1, f_2 \in C^{\infty}({\mathcal M})$ and 
$f_3 = \{ f_1, f_2 \}$, Poisson bracket of the two induced by the symplectic form,  then $[\hat{f}_1, \hat{f}_2] = -i \hbar \hat{f}_3$, ~\cite{Wo}. 

The motivating example  in our context would be the geometric
quantization of the  moduli space of flat connections on a principal 
$G$-bundle $P$ on a compact Riemann surface $\Sigma$, ~\cite{Wi}, ~\cite{ADW}. 
Let ${\mathcal A}$ be 
the space of Lie-algebra valued connections on the principal bundle $P$. Let 
${\mathcal N}$ be the moduli space of flat connections (i.e. the space of 
flat connections modulo the gauge group).  
One can construct   the determinant line bundle of the Cauchy-Riemann 
operator, namely, ${\mathcal L}
= \wedge^{\rm{top}} ({\rm Ker} \bar{\partial}_A)^{*} \otimes \wedge^{\rm{top}}({\rm Coker}
\bar{\partial}_A)$ on ${\mathcal A}$, ~\cite{Q}. The curvature induced by the Quillen metric on this bundle coincides with the natural K\"{a}hler form 
${\mathcal A}$ namely, 
$-{\rm Tr} \int_{\Sigma} \alpha \wedge \beta$, 
where $\alpha, \beta \in T_{A} {\mathcal A} = \Omega^{1}(M, {\rm ad} P).$
It can be shown, using a moment map construction, that this symplectic form
descends to the moduli space of flat connections ${\mathcal N}$.
The determinant line bundle is also well-defined on ${\mathcal N}$ and is the 
candidate for the prequantum line bundle of the geometric quantization. 

Inspired by this construction, we constructed three prequantum line bundles on 
the moduli space of  solutions to 
the self-duality equations over a Riemann surface ~\cite{D}, ~\cite{D1}
corresponding to the three symplectic forms which give rise to the 
hyperK\"{a}hler structure of the moduli space. 

In this paper we geometrically quantize the moduli space of vortex equations.
Geometric quantization of the  vortex moduli space  has been done before in 
~\cite{BR} and ~\cite{R}.
In the first paper the authors use algebraic geometry and in the second paper, 
the author uses the special form of the moduli space when the Riemann surface 
is a sphere. It would be interesting see what is the relation of the present
quantization to the ones in ~\cite{BR} and ~\cite{R}. The relation maynot be 
straightforward since  in the present quantization we find a whole family 
of (topologically equivalent, but perhaps holomorphically non-equivalent) 
prequantum line bundles ${\mathcal P}_{\Psi_0}$ whose curvatures 
correspond to a family 
of symplectic forms $\Omega_{\Psi_0}$ parametrised by $\Psi_0$ a section of a 
line bundle on the Riemann surface, as explained later. This symplectic form 
$\Omega_{\Psi_0}$ is a variant of the standard symplectic form $\Omega$ on the 
vortex moduli space. 

The vortex equations are as follows. Let $M$ be a compact Riemann 
surface and let $\omega = h^2 dz \wedge d \bar{z}$ be the purely imaginary 
volume form on it, (i.e. $h$ is real). 
Let  $A$ be a unitary connection on a principal $U(1)$ bundle $P$
 i.e. $A$ is a purely imaginary valued one form i.e. $A = A^{(1,0)} + A^{(0,1)}$ such that $A^{(1,0)} = -\overline{A^{(0,1)}}$.   
Let $L$ be a complex line bundle associated to $P$ by the defining 
representation. 
  Let $\Psi$ be a section of $L$, i.e. 
$\Psi \in  \Gamma(M,L)$ and $\bar{\Psi}$ be a section of its dual, $\bar{L}$. 
There is a Hermitian metric $H$ on $L$, i.e. the inner product $<\Psi_1, \Psi_2>_H = \Psi_1 H \bar{\Psi}_2$ is a smooth function on $M$. (Here $H$ is real). 

The pair $(A, \Psi)$ will be said to satisfy the vortex equations if

\hspace{1in} $(1)$ $ \rm{\;\;\;\;\;}$ $ F(A) = (1-|\Psi|^2_H) \omega,$

\hspace{1in} $(2)$ $\rm{\;\;\;\;\;}$ $\bar{\partial}_A \Psi = 0,$

where $F(A)$ is the curvature of the connection $A$ and $d_A = \partial_A + \bar{\partial}_A $ is the decomposition of the covariant derivative operator
into $(1,0)$ and $(0,1)$ pieces.  
Let ${\mathcal S}$ be the space of solutions to $(1)$ and $(2)$.
 There is a gauge group $G$ acting on the space of $(A, \Psi)$ which leaves the equations invariant. We take the group $G$ to be abelian and locally it looks like ${\rm Maps} (M, U(1)).$ If $g$ is an $U(1)$ gauge transformation then 
$(A_1, \Psi_1)$ and $(A_2, \Psi_2)$ are gauge equivalent if 
$A_2 = g^{-1}dg + A_1 $ and $\Psi_2 = g^{-1} \Psi_1$. 
  Taking the quotient by the gauge group of ${\mathcal S}$ gives  the moduli 
space of solutions to these 
equations and is denoted by ${\mathcal M}$.
It is well known that   there is a natural metric on the moduli space 
${\mathcal M}$ and in fact the metric is K\"{a}hler, see ~\cite{S}, ~\cite{MN},
~\cite{M}, ~\cite{R}, ~\cite{Ba}, ~\cite{CM}, ~\cite{BaM} and the references there.

In this paper, we show the metric explicitly and write down the symplectic 
(in fact, the K\"{a}hler form ) $\Omega$ arising from this metric and the 
complex structure. This is because 
some modification of this symplectic form gives us a whole family of 
symplectic forms $\Omega_{\Psi_0}$ parametrised by a fixed section 
$\Psi_0$ of the line bundle $L$ which vanishes on a set of measure zero. 
$\Omega_{\Psi_0}$ coincide with $\Omega$ when 
$L$ is a trivial bundle with $|\Psi_0|_H = 1.$  In fact $\Omega_{\Psi_0}$ is a 
K\"{a}hler form on the moduli space. We show that there exists a holomorphic 
prequantum line bundle, namely, a determinant line bundle, whose Quillen 
curvature is proportional to the sympletic form $\Omega_{\Psi_0}.$
Thus as $\Psi_0$ varies, we get a whole family of prequantum line bundles 
which are topologically equivalent, but perhaps not holomorphically 
equivalent.

\section{Metric and symplectc forms}

Let ${\mathcal A} $ be the space of all unitary connections on $P$ and 
$\Gamma (M, L)$ be sections of $L$. 
Let ${\mathcal C} = {\mathcal A} \times \Gamma (M, L)$ be the configuration 
space on which 
equations $(1)$ and $(2)$ are imposed. Let $p= (A, \Psi) \in {\mathcal C}$, $X
= ( \alpha_1, \beta)$, $Y= (\alpha_2, \eta)$
$\in T_p {\mathcal C} \equiv \Omega^1(M, i \RR) \times \Gamma(M, L) $ i.e.
 $\alpha_i = \alpha_i^{(0,1)} + \alpha_i^{(1,0)}$ such that 
$\overline{\alpha_i^{(0,1)}}= - \alpha_i^{(1,0)}, i = 1,2.$
On ${\mathcal C}$ one can define a metric
\begin{eqnarray*}
 {\mathcal G} ( X, Y) = \int_M *_1\alpha_1 \wedge \alpha_2  + 2i \int_M Re < \beta, \eta>_H \omega\end{eqnarray*}
and an almost complex structure  ${\mathcal I} = \left[
\begin{array}{cc}
*_1 & 0  \\
0 & i 
\end{array} \right] : T_p {\mathcal C} \rightarrow T_p {\mathcal C}$
where   $*_1: \Omega^{1} \rightarrow \Omega^{1}$ is  the Hodge star
operator on $M$ such that $*_1 (\eta dz) = - i \eta dz$ and $*_1 (\bar{\eta} d \bar{z}) = i \bar{\eta} d \bar{z}$. 

It is easy to check that ${\mathcal G} $ is positive definite. In fact, if  $\alpha_1 = \alpha^{(1,0)} + \alpha^{(0,1)} = a dz - \bar{a} d\bar{z} $ is an 
imaginary valued $1$-form,  $*_1 \alpha_1 = -i (adz + \bar{a} d \bar{z})$ and
${\mathcal G}(X, X) = 4 \int_M |a|^2 dx \wedge dy  + 4 \int_M |\beta|^2_H h^2 dx \wedge dy$ 
where $\omega = h^2 dz \wedge d \bar{z} = -2i h^2 dx \wedge dy$.  

{\bf The symplectic form $\Omega$}

We define
\begin{eqnarray*}
\Omega(X, Y) &=& -\int_{M} \alpha_1 \wedge \alpha_2 + 2i \int_{M} Re <
i \beta , \eta>_H \omega\\ 
&=& -\int_{M} \alpha_1 \wedge \alpha_2 - \int_{M}( \beta H \bar{\eta} - \bar{\beta} H \eta ) \omega
\end{eqnarray*}
such that $ {\mathcal G} ({\mathcal I} X, Y) = \Omega ( X, Y).$
Moreover, we have the following:

\begin{proposition} The metrics ${\mathcal G}$,  the symplectic form $\Omega$,
and the almost complex structure  ${\mathcal I}$ are invariant
under the gauge group action on ${\mathcal C}$. \label{inv}
\end{proposition}

\begin{proof}
 Let $p = (A, \Psi) \in {\mathcal C}$ and $g \in G, $ the gauge group, where
$g \cdot p = (A + g^{-1} dg, g^{-1} \Psi)$.

Then $g_* : T_p {\mathcal C} \rightarrow T_ {g \cdot p} {\mathcal C} $ is given
by the mapping $(Id, g^{-1})$ and it is now easy to check that $g$ and
$\Omega$ are invariant and ${\mathcal I}$ commutes with $g_*$.
\end{proof}

\begin{proposition}
\label{props} The equation $(1)$  can be realised as a moment
map $\mu = 0$ with respect to the action of the gauge group and
the symplectic form $\Omega$. \label{moment1}
\end{proposition}

\begin{proof}
Let $\zeta \in \Omega(M, i{\RR })  $ be the Lie algebra of the
gauge group (the gauge group element being $g = e^{ \zeta}$ ); note that 
$\zeta$ is purely imaginary.  
It generates a vector field $X_{\zeta}$ on ${\mathcal C}$ as follows :
$$X_{\zeta} (A, \Psi) = (d \zeta, -\zeta \Psi) \in T_p
{\mathcal C}$$ where $ p = (A, \Psi) \in {\mathcal C}.$

We show next that $X_{\zeta}$ is Hamiltonian. Namely, define
$H_{\zeta} : {\mathcal C} \rightarrow {\CC} $ as follows: $$
H_{\zeta} (p) = \int_{M} \zeta \cdot (  F_{A} -( 1-|\Psi|_{H}^2) \omega). $$  
Then for $X = (\alpha, \beta) \in T_p {\mathcal C}$,
 \begin{eqnarray*}
 dH_{\zeta} ( X ) & = & \int_M \zeta d \alpha  + \int_M \zeta   
( \Psi H \bar{\beta} + \bar{\Psi} H \beta )  \omega    \\
 &= &- \int_M (d \zeta) \wedge \alpha  + 2i  \int_M Re  
(i (-\zeta \Psi) H \bar{\beta}) \omega  \\
 & = & \Omega ( X_{\zeta},  X ),
 \end{eqnarray*}
where we use that $\bar{\zeta} = - \zeta$.

 Thus we can define the moment map $ \mu : {\mathcal C} \rightarrow
 \Omega^2 ( M, i{\RR} )= {\mathcal G}^* $ ( the dual of the Lie
 algebra of the gauge group)  to be $$ \mu ( A, \Psi)
 \stackrel{\cdot}{=} (F(A) - ( 1-|\Psi|_{H}^2)  \omega). $$ Thus equation 
$(1)$ is $\mu = 0$.
 \end{proof}

 \begin{lemma}
 Let ${\mathcal S}$ be the solution spaces to equation $(1)$ and $(2)$,  $X \in
 $  $ T_p {\mathcal S}$. Then ${\mathcal I}X $ $\in T_p {\mathcal S}$
 if and only if $X$ is ${\mathcal G}$-orthogonal to the gauge orbit $ O_p = G
 \cdot p$. \label{ortho}
 \end{lemma}

 \begin{proof}
 Let $X_{\zeta} \in T_p O_p,$ where $\zeta $ $\in $ $\Omega^{0} (M,
i {\RR })$,
${\mathcal G}( X , X_{\zeta} ) = -\Omega ({\mathcal I} X, X_{\zeta} ) =
- \int_M \zeta \cdot d \mu ( {\mathcal I} X ),$ and therefore
${\mathcal I} X$ satisfies the linearization of equation $(1)$ iff
$ d \mu ({\mathcal I} X)  = 0$, i.e.,  iff ${\mathcal G} (X, X_{\zeta}) = 0$ for all
$\zeta$. Second, it is easy to check that ${\mathcal I} X$ satisfies the
 linearization of equation $(2)$ whenever $X$ does.
 \end{proof}

 \begin{theorem}
 ${\mathcal M} $  has a natural symplectic structure and an almost
 complex structure  compatible with the symplectic form $\Omega $
 and the metric ${\mathcal G}$. \label{alcom}
 \end{theorem}

 \begin{proof}
 First we show that the almost complex structure descends to
 ${\mathcal M}$. Then using this and the symplectic quotient
 construction we will show that $\Omega$ gives a symplectic
 structure on ${\mathcal M}$.

 (a) To show that ${\mathcal I}$ descends as  an almost complex
 structure we let $pr: {\mathcal S} \rightarrow {\mathcal S}/G =
 {\mathcal M}$ be the projection map and set $[p] = pr (p)$. Then
 we can naturally identify $T_{[p]}  {\mathcal M} $ with the
 quotient space $T_p {\mathcal S} / T_p O_p, $ where $ O_p = G
 \cdot p $ is the gauge orbit. Using the metric ${\mathcal G}$ on ${\mathcal
 S}$ we can realize $T_{[p]} {\mathcal M}$ as a subspace in $T_p
 {\mathcal S},$ ${\mathcal G}$-orthogonal to  $T_p O_p$. Then by lemma
 ~\ref{ortho}, this subspace is invariant under ${\mathcal I}$.
Thus $I_{[p]} ={\mathcal I} |_{T_p (O_p )^{\perp}}$, gives the desired
almost  complex structure. This construction does not depend on the choice of
$p$ since ${\mathcal I}$ is $G$-invariant.

 (b) The symplectic structure $\Omega$ descends to $\mu^{-1}(0) /
 G$, (by proposition ~\ref{props} and by the Marsden-Weinstein
 symplectic quotient construction,~\cite{GS}, ~\cite{H}, since the
 leaves of the characteristic foliation are the gauge orbits). Now,
 as a $2$-form $\Omega$ descends to ${\mathcal M}$,   due to
 proposition (~\ref{inv}) so does the metric ${\mathcal G}$. 
 Closure of $\Omega$ is easy. We check that
 equation $(2)$ does not give rise to new degeneracy of
 $\Omega$ (i.e. the only degeneracy of $\Omega$ is due to $(1)$
 but along gauge orbits). Thus $\Omega $ is symplectic on ${\mathcal M}$.
 Since ${\mathcal G}$ and ${\mathcal I}$ descend to ${\mathcal M}$ the latter 
is symplectic and almost complex.
 \end{proof}

{\bf The family of symplectic forms $\Omega_{\Psi_0}$ }

Choose a fixed ${\Psi_0} \in \Gamma (M, L) $ such that $|\Psi_0|_H = 0$ only 
on a set of measure zero on $M$. (This $\Psi_0$ has nothing to do with $\Psi$). 

Define a symplectic form on ${\mathcal C}$ as
\begin{eqnarray*}
\Omega_{\Psi_0}(X, Y) &=& -\int_{M} \alpha_1 \wedge \alpha_2 + 2i \int_{M} Re <
i \beta , \eta>_H  |\Psi_0|_H^2 \omega \\
&=& -\int_{M} \alpha_1 \wedge \alpha_2 - \int_{M}( \beta H \bar{\eta} - \bar{\beta} H \eta ) |\Psi_0|^2_H \omega
\end{eqnarray*}

$|\Psi_0|^2_H$ plays the role of  a conformal rescaling of the volume form 
$\omega$ on $M$ 
which appears in $\Omega$, where we allow the conformal factor to have zeroes 
on sets of measure zero. 

\begin{theorem}
$\Omega_{\Psi_0}$ descends to ${\mathcal M}$ as a symplectic form.
\end{theorem}

\begin{proof}
Let $p= (A, \Psi).$

It is easy to show that $\Omega_{\Psi_0}$ is closed (this follows from the fact that on ${\mathcal C}$ it is a constant form -- does not depend on $(A, \Psi)$). 
We have to show it is 
non-degenerate. 

Suppose there exists $(\alpha_1, \beta) \in T_{[p]}({\mathcal M})$
s.t. $$\Omega_{\Psi_0} ((\alpha_2, \eta), (\alpha_1, \beta)) = 0$$ 
$\forall$ $(\alpha_2, \eta) \in T_{[p]} ({\mathcal M})$.  
Using the metric ${\mathcal G}$  we identify $T_{[p]} {\mathcal M}$ with 
the subspace in $T_p {\mathcal S},$  ${\mathcal G}$-orthogonal to  
$T_p O_p$ (i.e. the tangent space to the moduli space is identified to the tangent space to solutions which are orthogonal to the gauge orbits, the orthogonality is 
with respect to the metric ${\mathcal G}$.)
Thus $(\alpha_1, \beta), (\alpha_2, \eta)$ satisfy the linearization of equation $(1)$ and $(2)$ and ${\mathcal G} ((\alpha_1, \beta), X_{\zeta}) = 0 $ and 
 ${\mathcal G} ((\alpha_2, \eta), X_{\zeta}) = 0$ for all $\zeta$. 

Now, by ~\ref{ortho}, ${\mathcal I} (\alpha_2, \eta) \in T_{p} S.$ Also, 
\begin{eqnarray*}
{\mathcal G}({\mathcal I} (\alpha_1, \beta), X_{\zeta}) &=&  
\Omega ((\alpha_1, \beta), X_{\zeta}) \\ 
&=& -\int_M \zeta  d \mu ((\alpha_1, \beta))\\
&=& 0
\end{eqnarray*}
since $d \mu ((\alpha_1, \beta))= 0$ is precisely one of the equations
saying  that $(\alpha_1, \beta) \in T_p S$.
Thus ${\mathcal I} (\alpha_1, \beta) \in T_{[p]} {\mathcal M},$ 
(since it is in $T_p S$ and ${\mathcal G}$-orthogonal to gauge orbits).

Take $ (\alpha_2, \eta) = {\mathcal I} (\alpha_1, \beta) = (*_1 \alpha_1, i \beta). $ Then
\begin{eqnarray*}
0 &=&  \Omega_{\Psi_0} ({\mathcal I} (\alpha_1, \beta),  (\alpha_1, \beta)) \\ 
&=& -\int_M ( *_1 \alpha_1 \wedge \alpha_1) + 2i \int_M Re < i(i \beta),  \beta>_H |\Psi_0|_H^2 \omega \\
&=& -4\int_M  |a|^2 dx \wedge dy  - 4\int_M | \beta|_H^2 |\Psi_0|_H^2  h^2 dx \wedge dy
 \end{eqnarray*}
where $\omega = -2i h^2 dx \wedge dy $ and 
$\alpha_1 = a dz - \bar{a} d \bar{z}  \in \Omega^1(M, i \RR)$ 
and $*_1\alpha_1 = -i( a dz + \bar{a} d \bar{z} )$. 
By negativity of both the terms and the fact that $\Psi_0$ has zero on a set of measure zero on $M$,   $(\alpha_1, \beta) = 0$ a.e. 
Thus $\Omega_{\Psi_0} $ is symplectic. 
\end{proof}

\section{Prequantum line bundle}
 In this section we briefly review the Quillen construction of the determinant 
line bundle of the Cauchy Riemann operator  $\bar{\partial}_A = \bar{\partial} + A^{(0,1)}$, ~\cite{Q},
which  enables  us to construct prequantum line bundle on the vortex moduli 
space.

First let us note that a connection $A$ on a $U(1)$-principal bundle induces 
a  connection on any associated line bundle $L$. 
We will denote this connection also by $A$ since  the same ``
Lie-algebra valued $1$-form'' $A$ (modulo representations)  gives  a covariant 
derivative operator enabling you to take derivatives of  sections of $L$ 
~\cite{N}, page 348.
A very clear description of
the determinant line bundle can be found in ~\cite{Q} and
~\cite{BF}. Here we mention  the formula for the Quillen curvature
of the determinant line bundle $\wedge^{\rm top} (Ker \bar{\partial}_A)^{*}
\otimes \wedge^{\rm top}(Coker \bar{\partial}_A) = {\rm det}(\bar{\partial}_A)$,
 given the canonical unitary connection $\nabla_Q$, induced by the Quillen 
metric,~\cite{Q}.
Recall that the affine space ${\mathcal A}$ (notation as
in ~\cite{Q}) is an infinite-dimensional K\"{a}hler manifold. Here
each connection  is identified with its $(0,1)$ part which is the holomorphic 
part. Since the connection $A$ is unitary (i.e. $A = A^{(1,0)} + A^{(0,1)}$  
s.t. $\overline{A^{(1,0)}} = -A^{(0,1)}$) this identification is easy.
 In fact, for every $A \in {\mathcal A}$,
$T_A^{\prime} ({\mathcal A}) \stackrel{~}{=} \Omega^{0,1} (M, i \RR)$ and 
the corresponding K\"{a}hler  form  is given by 
\begin{eqnarray*}
F(\alpha_1^{(0,1)}, \alpha_2^{(0,1)}) &=&  {\rm Re} \int_M  (\alpha_1^{(0,1)} \wedge *_2 \alpha_2^{(0,1)}),\\
&=& - \frac{1}{2} \int_M \alpha_1 \wedge \alpha_2
\end{eqnarray*}
where $\alpha^{(0,1)}, \beta^{(0,1)} \in \Omega^{0,1} (M, i\RR)$  and 
$ *_2 $ is the Hodge-star operator such that 

$*_2(\eta dz) = - \bar{\eta} d\bar{z}$ and $*_2 (\bar{\eta} d \bar{z}) =  \eta d z$ and  we have used
$\overline{\alpha_i^{(0,1)}} = - \alpha_i^{(1,0)}$, $i =1,2$. 
Let $\nabla_Q$ be the conection induced from the Quillen metric. Then the 
Quillen curvature of ${\rm det} (\bar{\partial}_A)$ is 
$${\mathcal F}(\nabla_Q) = \frac{i}{ \pi} F.$$

\section{Prequantum bundle on ${\mathcal M}$}

First we note that to the connection $A$ we can add any one form and
still obtain a derivative operator. 

Let $\omega = h^2 dz \wedge d \bar{z}$ where recall $h$ is real. 
Let $\theta = h dz$ , $\bar{\theta} = h d \bar{z} $ be 1-forms (~\cite{GH}, page 28)
 such that $ \omega=\theta \wedge \bar{\theta} = h^2 dz \wedge d \bar{z}$.
Let $\Psi_0$ be the same $fixed$ section used to define $\Omega_{\Psi_0}$. Recall
$\Psi_0$ has zero on a set of measure zero on $M$.
Note $\Psi H \bar{\Psi}_0$ is a smooth function on $M$.
Thus  $B^{(0,1)}= \Psi H \bar{\Psi}_0 \bar{\theta} $ is a $(0,1)$-form we would like
to add to the connection $A^{(0,1)}$ to make another connection form. 
Note that $B^{(0,1)}$ is gauge invariant, since $\Psi$ and $\Psi_0$ gauge 
transform in the same way.  Note that $A^{(0,1)} \pm B^{(0,1)}$ are the $(0,1)$
parts of  a connection defined by $A \pm B = A^{(0,1)} \pm B^{(0,1)} + 
A^{(1,0)} \pm B^{(1,0)}$ where $B^{(1,0)}$ is defined to be 
$\overline{B^{(1,0)}} = -B^{(0,1)}$. 

{\bf Definitions:} Let us denote by ${\mathcal L}_{\pm} = 
{\rm det} (\bar{\partial}+ A^{(0,1)} \pm B^{(0,1)})$ a determinant bundle on ${\mathcal J}_{\pm} = \{ A^{(0,1)} \pm  \Psi H \bar{\Psi}_0 \bar{\theta}| A \in {\mathcal A}, \Psi \in \Gamma(M, L) \}$ which is isomorphic to 
${\mathcal C} = {\mathcal A} \times \Gamma(M,L)$. 

Thus ${\mathcal P}_{\Psi_0} = {\mathcal L}_{+} \otimes {\mathcal L}_{-}$
well-defined line bundle on ${\mathcal C}$.

\begin{lemma}
${\mathcal P}_{\Psi_0}$ is a well-defined line bundle over 
${\mathcal M} \subset {\mathcal C}/G$, 
where $G$ is the gauge group.
\end{lemma}
\begin{proof}
First consider the Cauchy-Riemann operator 
$ D= \bar{\partial} + A^{(0,1)} + B^{(0,1)}$. Under gauge transformation 
$D=\bar{\partial} + A^{(0,1)}  + B^{(0,1)}
\rightarrow D_g= g(\bar{\partial} + A^{(0,1)} + B^{(0,1)})g^{-1} $.
We can show that the operators $D$ and $D_g$ have isomorphic 
kernel and cokernel and their corresponding Laplacians have the 
same spectrum and the eigenspaces are of the same dimension. Let 
$\Delta$ denote the Laplacian corresponding to $D$ and $\Delta_g$ 
that corresponding to $D_g$. 
The Laplacian is $\Delta = \tilde{D} D$ where 
$\tilde{D} = \partial + A^{(1,0)} + B^{(1,0)}$, where recall $\overline{A^{(1,0)}} = -A^{(0,1)}$ and $\overline{B^{(1,0)}} = - B^{(0,1)}$. Note that  
$\tilde{D} \rightarrow  \tilde{D}_g = g \tilde{D} g^{-1}$ under gauge transformation. Then $\Delta_g = g \Delta g^{-1}$. 
Thus the isomorphism of eigenspaces is  $s \rightarrow g s$. We describe here how to define the line bundle on the moduli space.
Let $K^a(\Delta)$ be the direct sum of 
eigenspaces of the operator $\Delta$ of 
eigenvalues $< a$, over the open subset 
$U^a = \{A^{(0,1)} + B^{(0,1)} | a \notin {\rm Spec} \Delta \}$ of the affine 
space ${\mathcal J_{+}}.$ The determinant line bundle is defined using the exact sequence
$$ 0 \rightarrow {\rm Ker} D \rightarrow K^a(\Delta) \rightarrow 
D(K^a(\Delta)) \rightarrow {\rm Coker} D \rightarrow 0$$ 
Thus 
one identifies 

$\wedge^{{\rm top} }({\rm Ker} D)^* \otimes \wedge^{{\rm top} }
({\rm Coker} D)$ with 
 $\wedge^{{\rm top}}(K^a(\Delta))^* \otimes \wedge^{{\rm top}} 
(D(K^a(\Delta)))$  (see ~\cite{BF},  for more details) and 
there is an isomorphism of the fibers as $D \rightarrow D_g$. 
Thus one can identify 

$$ \wedge^{{\rm top}}(K^a(\Delta))^* \otimes \wedge^{{\rm top}} 
(D(K^{a}(\Delta))) \equiv
\wedge^{{\rm top}}(K^a(\Delta_g))^* \otimes \wedge^{{\rm top}} 
(D(K^{a}(\Delta_g))).$$
By extending this definition from 
$U^a$ to $V^a = \{(A, \Psi)| a \notin {\rm Spec} \Delta \}$, 
an open subset of ${\mathcal C}$,  we can define the fiber over 
the quotient space $V^a/G$ to be the 
equivalence class of this fiber. Covering ${\mathcal C}$ with open sets of the 
type $V^a$, we can define it on ${\mathcal C}/G$. Then we can restrict it to
${\mathcal M} \subset {\mathcal C}/G$.

Similarly one can deal with the other case of  
$\bar{\partial} + A^{(0,1)} - B^{(0,1)}$.
Let $([A], [\Psi]) \in {\mathcal C}/G,$ 
where $[A], [\Psi]$ are gauge equivalence classes of $A, \Psi$, 
respectively.  Then associated to the equivalence class $([A], [\Psi])$ in the 
base space, there is an 
equivalence class of fibers coming from the identifications 
of ${\rm det} (\bar{\partial} + A^{(0,1)} - B^{(0,1)})$ with ${\rm det}(g(\bar{\partial} + A^{(0,1)} - B^{(0,1)})g^{-1})$ as mentioned in the previous case. 

 This way one can prove that  ${\mathcal P}_{\Psi_0}$ is well defined 
on ${\mathcal C}/G$. Then we restrict it to 
${\mathcal M} \subset {\mathcal C}/G$.
\end{proof}

{\bf Curvature and symplectic form:}

Let $p = (A, \Psi) \in S$. Let $X, Y \in T_{[p]}{\mathcal M}$. 
Since $T_{[p]}{\mathcal M}$ can be identified with a subspace in 
$T_p S$ orthogonal to $T_p O_p$, if we write 
$X =(\alpha_1, \beta)$ and $Y=(\alpha_2, \eta)$, 
$\alpha_1, \alpha_2 \in T_{A} {\mathcal A} = \Omega^{1}(M, i \RR),$  and 
$\beta, \eta \in T_{\Psi} \Gamma(M,L) = \Gamma(M, L)$, 
 then 
$X,Y$ can be said to satisfy a) $X, Y \in T_p S$ and b) 
$X, Y$ are ${\mathcal G}$-orthogonal to $T_p O_p $, the tangent space to the gauge orbit.  

Let ${\mathcal F}_{{\mathcal L}_{\pm}}$
 denote the Quillen curvatures of the
determinant line bundles ${\mathcal L}_{\pm}$,  
 respectively. ${\mathcal L}_{\pm}$ are determinants  of Cauchy-Riemann operators of the connections $  A^{(0,1)}\pm \Psi H \bar{\Psi}_0 \bar{\theta}.$ Thus in the curvature, we will have $\alpha_{1}^{(0,1)} \pm \beta H \bar{\Psi}_{0} \bar{\theta}$ and $\alpha_2^{(0,1)} \pm \eta H \bar{\Psi}_0  \bar{\theta},$ 
(see Quillen's formula in the section above). 
\begin{eqnarray*}
{\mathcal F}_{{\mathcal L}_{\pm}} (X,Y)&=& 
\frac{i}{\pi} {\rm Re} \int_M (\alpha_{1}^{(0,1)} \pm \beta H \bar{\Psi}_{0} \bar{\theta}) \wedge *_2(\alpha_2^{(0,1)} \pm \eta H \bar{\Psi}_0  \bar{\theta}) \\
&=& \frac{i}{ \pi}{\rm Re} \int_M (\alpha_{1}^{(0,1)} \pm \beta H \bar{\Psi}_{0} \bar{\theta}) \wedge (- \alpha_2^{(1,0)} \pm \bar{\eta} H \Psi_0 \theta ) 
\end{eqnarray*}

Note that ${\rm Re} \int_M \alpha_1^{(0,1)} \wedge \alpha_2^{(1,0)} = \frac{1}{2} \int_M \alpha_1 \wedge \alpha_2 $  
where we have used the fact that $\alpha_i = \alpha^{(0,1)} + \alpha_i^{(1,0)}$ s.t. $\overline{\alpha_i^{(0,1)}} = -\alpha_i^{(1,0)}$, $i=1,2$,
We have also used that $\bar{\theta} \wedge \theta = -\omega = 2i h^2 dx \wedge dy,$ is purely imaginary. 
One can easily compute that 
\begin{eqnarray*}
& &({\mathcal F}_{{\mathcal L}_{+}} + {\mathcal F}_{{\mathcal L}_{-}})(X,Y)  \\
&=& \frac{i}{ \pi}[-\int_M \alpha_{1} \wedge \alpha_2 -  \int_M (\beta H \bar{\eta} - \bar{\beta} H \eta)  |\Psi_0|_H^2 \omega )] \\
&=& \frac{i}{\pi}\Omega_{\Psi_0}(X,Y)
\end{eqnarray*}

Now  $A^{(0,1)}$ is holomorphic w.r.t. the complex structure 
$*_1$  and $\Psi$ is holomorphic w.r.t. multiplying by $i$, 
$A^{(0,1)} \pm B^{(0,1)}$ is holomorphic w.r.t. the complex structure
${\mathcal I}$. Thus ${\mathcal L}_{+},$ ${\mathcal L}_{-}$ and 
${\mathcal P}_{\Psi_0}$ are holomorphic, (same argument as in ~\cite{Q}).

Thus, we have proven the following theorem:
\begin{theorem}
$ {\mathcal P}_{\Psi_0} = {\mathcal L}_{+} \otimes {\mathcal L}_{-}$ is a well-defined
holomorphic line bundle on ${\mathcal M}$ whose Quillen curvature is
${\mathcal F}_{{\mathcal L}_{+}} + {\mathcal F}_{{\mathcal L}_{-}} $
which is $\frac{i}{\pi}\Omega_{\Psi_0}$. Thus ${\mathcal P}_{\Psi_0}$ is a 
prequantum bundle on ${\mathcal M}$.
\end{theorem}

{\bf Polarization:} In passing from prequantization to quantization, one needs 
a polarization. It can be shown that the almost complex structure 
${\mathcal I}$ is integrable on ${\mathcal M}$ , (see, for example, Ruback's 
argument mentioned in ~\cite{S} or matscinet review of  
~\cite{M}, ~\cite{MN}).  In fact, $\Omega_{\Psi_0}$ is a K\"{a}hler form and 
${\mathcal G}_{\Psi_0} (X,Y) = \Omega_{\Psi_0}(X, {\mathcal I}Y)$
is a K\"{a}hler metric on the moduli space (since it is positive definite).
 ${\mathcal P}_{\Psi_0}$ is a holomorphic line bundle on ${\mathcal M}$. 
Thus we can take holomorphic square integrable sections of 
${\mathcal P}_{\Psi_0}$ as our Hilbert space.
The dimension of the Hilbert space is not easy to compute. (For instance,
the holomorphic sections of the determinant line bundle on the moduli space of  flat connections for $SU(2)$ gauge group is the Verlinde dimension of the 
space of  conformal blocks in a 
certain conformal field theory). This would be a topic for future work.

{\bf Remark:}

As $\Psi_0$ varies, the corresponding line bundles are  all topologically
equivalent since the curvature forms have to be of integral cohomology
and that would be constant. Thus they  have the same Chern class. However they 
maynot be holomorphically equivalent.

{\bf Acknowledgement}
I would like to thank Professor Jonathan Weitsman for pointing out the 
equivalence of these line bundles.

School of Mathematics, Harish Chandra Research Institute, Jhusi,
Allahabad, 211019, India. email: rkmn@mri.ernet.in

\end{document}